\documentclass{cimento}

\usepackage{graphicx}  

\newcommand{\beq}{\begin{equation}}
\newcommand{\eeq}{\end{equation}}
\newcommand{\bea}{\begin{eqnarray}}
\newcommand{\eea}{\end{eqnarray}}

\def\laq{~\raise 0.4ex\hbox{$<$}\kern 
-0.8em\lower 0.62ex\hbox{$\sim$}~}
\def\gaq{~\raise 0.4ex\hbox{$>$}\kern 
-0.7em\lower 0.62ex\hbox{$\sim$}~}

\def \pa {\partial}
\def \ti {\widetilde}

\def \ra {\rightarrow}

\def \da {\delta}
\def \Da {\Delta}
\def \b {\beta}
\def \a {\alpha}

\def \ga {\gamma}
\def \sg {\sigma}

\def \r {\rho}
\def \om {\omega}
\def \Om {\Omega}

\def \ls {\lambda_{\rm s}}
\def \lp {\lambda_{\rm P}}
\def \Ms {M_{\rm s}}
\def \Mp {M_{\rm P}}

\def \fpu {\dot{\phi}}
\def \fpp {\ddot{\phi}}
\def \fb {\overline \phi}
\def \fbp {\dot{\fb}}
\def \fbpp {\ddot{\fb}}
\def \rb {\overline \rho}
\def \pb {\overline p}

\begin{document}

\thispagestyle{empty}

\begin{flushright}
BA-TH/07-651\\
CERN-PH-TH/2007-026\\
hep-th/0703055
\end{flushright}

\vspace{0.3 cm}

\begin{center}

\huge{String Theory and\\ Pre-big bang Cosmology}

\vspace{0.7cm}

\large{M. Gasperini$^{1,2}$ and G. Veneziano$^{3,4}$}

\vspace{0.5cm}
\normalsize

\smallskip
$^{1}${\sl Dipartimento di Fisica, Universit\`a di Bari, \\
Via G. Amendola 173, 70126 Bari, Italy}

\smallskip
$^{2}${\sl Istituto Nazionale di Fisica Nucleare, Sezione di Bari, 
Bari, Italy}

\smallskip
$^{3}${\sl CERN, Theory Unit, Physics Department, \\ CH-1211 Geneva 23, Switzerland} 

\smallskip
$^{4}${\sl Coll\`ege de France, 11 Place M. Berthelot, 75005 Paris, France} 

\vspace{0.8cm}

\begin{abstract}
In string theory, the traditional picture of a Universe that emerges from the inflation of a very small and highly curved space-time patch is a possibility, not a  necessity: quite different initial conditions are possible, and not necessarily unlikely. In particular, the duality symmetries of string theory suggest scenarios in which the Universe starts inflating from an initial state characterized by very small curvature and interactions. Such a state, being gravitationally unstable, will evolve towards higher curvature and coupling, until string-size effects and loop corrections make  the Universe ``bounce" into a standard, decreasing-curvature regime.  In such a context, the hot big bang of conventional cosmology is replaced by  a ``hot big bounce" in which the bouncing and heating mechanisms originate from the quantum production of particles in the high-curvature, large-coupling pre-bounce phase. Here we briefly summarize the main features of this inflationary scenario, proposed a quarter century ago. In its simplest version (where it represents an alternative and not a complement to standard slow-roll inflation) it can produce a viable spectrum of density perturbations, together with a tensor component characterized by a ``blue" spectral index with a peak in the GHz  frequency range. That means, phenomenologically, a very small  contribution to a primordial B-mode in the CMB polarization, and the possibility of a large enough stochastic background of gravitational waves to be measurable by present or future gravitational wave detectors.
\end{abstract}
\end{center}

\begin{center}

Contribution to the special issue of IL NUOVO CIMENTO \\
\smallskip
published in honor of {Gaetano Vilasi} on the occasion of his $70$th birthday.\\
\bigskip
({\em Il Nuovo Cimento C}, Italian Physical Society, 2015)
\end{center}
\newpage

\title{String theory and pre-big bang cosmology\thanks{This paper is dedicated to our colleague and friend Gaetano Vilasi on the occasion of his $70$th birthday.}}
\author{M. Gasperini\from{ins:x} \ETC
 ~\atque
G. Veneziano\from{ins:y}}
\instlist{\inst{ins:x} Dipartimento di Fisica, Universit\`a di Bari, 
Via G. Amendola 173, 70126 Bari, Italy, \\
and INFN, Sezione di Bari - Bari, Italy
\inst{ins:y}CERN, Theory Unit, Physics Department, CH-1211 Geneva 23, Switzerland, \\
and Coll\`ege de France, 11 Place M. Berthelot, 75005 Paris, France}


\PACSes{\PACSit{04.60.Cf}
{98.80.Cq} -- {11.25.-w}}


\maketitle

\begin{abstract}
In string theory, the traditional picture of a Universe that emerges from the inflation of a very small and highly curved space-time patch is a possibility, not a  necessity: quite different initial conditions are possible, and not necessarily unlikely. In particular, the duality symmetries of string theory suggest scenarios in which the Universe starts inflating from an initial state characterized by very small curvature and interactions. Such a state, being gravitationally unstable, will evolve towards higher curvature and coupling, until string-size effects and loop corrections make  the Universe ``bounce" into a standard, decreasing-curvature regime.  In such a context, the hot big bang of conventional cosmology is replaced by  a ``hot big bounce" in which the bouncing and heating mechanisms originate from the quantum production of particles in the high-curvature, large-coupling pre-bounce phase. Here we briefly summarize the main features of this inflationary scenario, proposed a quarter century ago. In its simplest version (where it represents an alternative and not a complement to standard slow-roll inflation) it can produce a viable spectrum of density perturbations, together with a tensor component characterized by a ``blue" spectral index with a peak in the GHz  frequency range. That means, phenomenologically, a very small  contribution to a primordial B-mode in the CMB polarization, and the possibility of a large enough stochastic background of gravitational waves to be measurable by present or future gravitational wave detectors.

\end{abstract}


\section*{Foreword}

This work is a contribution to the special issue of IL NUOVO CIMENTO published in honor of Gaetano Vilasi. We met Gaetano many years ago, and we have been attending very frequently (and with great pleasure) the Workshops on  Theoretical Physics that he has been organizing in Vietri for almost thirty years. We have not (yet) signed a paper together with Gaetano, but we have often exchanged opinions and carried on fruitful discussions on our common fields of interest, general relativity and gravitational theory. It is thus a pleasure for us to celebrate his scientific career, his many achievements, and to join all his friend in wishing him  many future years of a long, happy and productive scientific life!

\section{Introduction}
\label{sec1}

The standard cosmological model, formulated and brought to completion by various authors during the second half of the last century (see for instance \cite{1}), provides us with an excellent description of the various important stages of our past cosmological history (such as the radiation era, the nucleosynthesis, the recombination era, the epoch of matter domination, \dots). At early enough times, however, such a model is to be modified by the introduction of a ``non-standard" epoch of accelerated cosmic  evolution, called ``inflation", which is needed in order to solve the horizon, flatness and entropy problems \cite{3} implied by the extrapolation, back in time, of the present state according to the standard cosmological equations. 

The introduction of a sufficiently long inflationary phase not only ``explains" the otherwise unnatural initial conditions required by the subsequent ``standard" evolution, but also provides a natural (and very efficient) mechanism for amplifying the quantum fluctuations of the fundamental cosmic fields in their ground state. This mechanism produces inhomogeneous ``seeds" for structure formation and for the fluctuations of the temperature of the Cosmic Microwave Background (CMB) radiation (see e.g. \cite{3a}), in remarkable agreement with recent observations (see e.g. the updated result listed in \cite{5}, and the last data of the PLANCK satellite reported in \cite{4}). 

The most popular implementation of  the inflationary scenario is provided by a suitable version the so-called model of ``slow-roll" inflation \cite{6}, in which the cosmological evolution is dominated by the potential energy of a cosmic scalar field (the ``inflaton"). The simplest and most conventional version of such model, however, is affected by various difficulties of conceptual nature. First of all, the peculiar properties of the inflaton  (mass, couplings, potential energy, \dots)  prevent a simple identification of this field  in the context of known models of fundamental interactions: the inflaton field has to be introduced {\em ad hoc}, and its properties are the result of a suitable fine-tuning of the relevant parameters.

Another difficulty is associated with the kinematic properties of the phase of slow-roll inflation, implying that the spatial size of the Hubble horizon $cH^{-1}$, proportional to the causal event horizon, becomes smaller and smaller as we go back in time towards the beginning of the inflation era (here $c$ is the speed of light and $H$ is the so-called Hubble parameter). It follows, for conventional values of the inflation curvature scale (corresponding, typically, to the GUT energy scale), that the proper (time-dependent) wavelength of the fluctuations presently observed on a large scale (for instance, the temperature fluctuations of the CMB radiation), when rescaled down at the beginning of inflation, has to be smaller than the Planck length, $\lp=(8\pi G \hbar/c^3)^{1/2} \sim 10^{-33}$ cm, if inflation lasts long enough to solve the above mentioned problems (here $G$  and $\hbar$ are the Newton and Planck constant, respectively). As a consequence, the initial conditions on the fluctuations of the matter fields and of the geometry are to be imposed inside an energy range which is presently unexplored, and in which the extrapolation of standard physics is -- at least -- questionable. This is the so-called ``trans-Planckian" problem \cite{7}. 

In addition, if the Hubble horizon is decreasing as we go back in time, then its inverse, the spacetime curvature-scale $H/c$, is necessarily increasing, and the extrapolation of the model unavoidably leads to an initial singularity -- or, at least, to the quantum gravity regime $cH^{-1} \sim \lp$ where we have to deal not only with an unknown fluctuation  dynamics, but also with an unknown dynamics of the background itself. A proof of  the unavoidable presence of the singularity, in the context of  potential-driven inflationary cosmology based on the Einstein equations, is given in \cite{8,8a,8b}. 

All these problems, as well as other important problems of primordial cosmology (why is our Universe four-dimensional? why does a small vacuum energy density -- the so-called ``dark energy" -- seem to survive until today after inflation? \dots), should find a satisfactory solution in the context of a truly unified  theory of all fundamental interactions. The best candidate for such theory, at present, is  (in our opinion) string theory, which is  based on the assumption that the fundamental components of all matter and force fields existing in nature are one-dimensional extended objects, called  ``strings". These objects are characterized by their tension (i.e., energy per unit of length) $T$ and, when quantized, have a characteristic size $\ls$ given by:
\beq
\ls = \sqrt{c \hbar/T} .
\eeq

As it will discussed in the following section, the consistency of string theory requires the existence of extra spatial dimensions besides the three ones we are familiar with. In such a context, two possibilities arise. 
In the most conventional unification scenario, in which the extra spatial dimensions are of size comparable to the string length scale  \cite{9}, the parameter $\ls$ turns out to be extremely small, 
$\ls \sim 10 \,\lp$. As a result, all effects pertaining to the finiteness of the string size only come into play at energy scales $\Ms c^2 =\hbar c/ \ls$, so high to be well outside the reach of present (direct) high-energy experiments. In such a case the predictive power of the theory is low, as what can be tested by accelerator experiments are only the predictions of the so-called low-energy string effective action. Unfortunately, at the present stage of our knowledge, such an action can only be derived within  perturbative computational techniques, and its precise form is unknown. 

This discouraging conclusion can be avoided, at least in principle, if our Universe contains extra spatial dimensions  compactified on length scales which are small if compared to macroscopic standards, but are nevertheless ``large" with respect to the parameter $\ls$ of string theory \cite{10,10a}. In this case the energy scale $\Ms c^2$ may be lowered well below the Planck scale $\Mp c^2= \hbar c/\lp$, even by many orders of magnitude, thus possibly approaching the TeV scale and the energy range accessible to present (or near future) accelerator experiments. 

In any case, the energy scales typical of string theory (whether near to, or far from, the Planck scale) should have been  reached during the primordial evolution of our Universe: string theory can then be properly applied to cosmology, to ask if (and how) inflation is naturally predicted, in such a context. The hope is to obtain, on one hand, a solution to the open problems of the conventional inflationary scenario and, on the other hand, a possible phenomenological signature of string theory, according to the historical tradition teaching us that fundamental gravitational theories have always been confirmed by astrophysical observations (as, for instance,  Newton's and Einstein's theories of gravity). 

In this paper, after a short presentation of string theory in Sect. \ref{sec2}, and a brief discussion of inflationary string cosmology in Sect. \ref{sec3}, we shall concentrate in Sect. \ref{sec4} on the so-called ``pre-big bang" scenario \cite{11}, based on the scale-factor duality symmetry \cite{12} typical of string theory. It will be shown, in particular,  that the Universe (thanks to a fundamental string-theory field, the dilaton) may start inflating from an initial configuration characterized by a very small curvature and a very large Hubble horizon, thus avoiding the singularity and trans-Planckian problems, and yet satisfying the properties required for a successful  scenario.

\section{String theory: a few basic concepts}
\label{sec2}

String theory is, at present, the only serious candidate for a fully quantum (as well as finite and unified) theory of gravity and of all known gauge interactions. It is therefore the natural framework  for  formulating and discussing gravitational problems in which quantum effects are expected to be non-negligible or, even worse, dominant, as in the case in which  a spacetime singularity  is approached. No surprise, therefore, that natural arenas for string theory are the physics of quantum black holes (in particular the end-point of their evaporation) and that of the cosmological (big bang) singularity. Before discussing the possible applications of string theory, however, it seems appropriate to explain why this theory includes gravity, in a natural and compelling way, unlike in  the conventional models of gauge unification. 

\subsection{Quantization and gravity}

We should recall, to this purpose, that an extended one-dimensional object like a string corresponds, dynamically, to a constrained system. For a complete and consistent description of the string motion we must satisfy, in fact, not only the Euler-Lagrange equations, but also a set of dynamical constraints (at any point along the string trajectory), expressed by the vanishing of appropriate functionals. In addition, we have to impose the required boundary conditions, different for the case of open strings (with non-coincident ends) and closed strings  
(without free ends, like a loop). 

When the model is quantized, the coordinates and momenta of the string are promoted to operators satisfying canonical commutation relations. Also  the constraints are now represented by (the so-called Virasoro) operatos $L_m$. The Hilbert subspace containing the physical states of the system is then formed by (all and only) those states satisfying the constraints in a ``weak" sense, i.e. those annihilated by the application of the Virasoro operators $L_m$ with $m \ge0$, appropriately ordered and regularized  (see e.g. \cite{13,13a}). As it turns out, this  is sufficient to eliminate all the negative-norm (so called ``ghosts") states.  The  lowest order constraint ($L_0=0$ up to a subtraction constant) corresponds to imposing the mass-shell condition $p_\mu p^\mu=M^2c^2$ determining the allowed energy levels of the string spectrum (here $p_\mu p^\mu$ is the square of the energy-momentum vector, and $M$ the rest mass of the string). 

A quantized bosonic string, on the other hand, corresponds to a system of infinitely many harmonic  oscillators vibrating along the spatial directions orthogonal to the string itself. The various states of the discrete string spectrum can be obtained (as in the theory of the elementary quantum oscillator) by applying the appropriate number of creation operators (along different spatial directions, in general) to the lowest energy level. As a consequence, the physical string states  can be ordered as a tower of states of growing  (discrete) mass and angular-momentum eigenvalues. And here we find the ``miracle" connecting strings to gravitational interactions.

Looking at the subset of massless eigenstates, in fact, we find that the open string spectrum contains (even in the simplest case) a vector $A_\mu$ which is transverse 
(i.e., it has a vanishing divergence, $\pa^\mu A_\mu=0$), and which can  be associated with an Abelian interaction of vector type, like the electromagnetic interaction. The closed string spectrum, instead, contains -- besides a scalar  $\phi$, the dilaton, and a second-rank antisymmetric tensor $B_{\mu\nu}$,  -- a symmetric tensor field $h_{\mu\nu}$ which is transverse and traceless ($\pa^\nu h_{\mu\nu}=0= h_\mu\,^\mu$), and which has all the required physical properties of the  graviton. Thus, unified models of fundamental interactions based on strings must {\em necessarily} include a tensor interaction of gravitational type. 

\subsection{Supersymmetry}

But the virtues of string theory as basis for a unified model of all interactions are not limited to this result. We should recall, indeed, that the spectrum of the bosonic string that we have considered contains, even after the elimination of the ghost states, other states of  ``tachyonic" type (i.e., states with $M^2<0$). To avoid such states (sources of instabilities in a quantum  theory context) the model of bosonic string has to be ``supersymmetrized". The standard procedure is to associate to the  coordinates $X^\mu$, determining the position of the string in the external ``target" space in which the string is embedded, the fermionic partners  $\psi^\mu$, transforming  as  two-component Majorana spinors on the (two-dimensional) world-sheet spanned by the string motion, and, in the Neveu-Schwarz-Ramond formulation,  as  Lorentz vectors (with index $\mu$) in the  target space manifold. This leads us to the so-called superstring models, which can be consistently formulated (from a quantum point of view) only in five different versions \cite{13,13a}. 

This generalization of the string model not only eliminates tachyons from the physical spectrum -- in a supersymmetric theory, the lowest allowed energy level corresponds to $M^2=0$, and negative values are forbidden -- but also automatically introduces in the model the fermions required to describe the fundamental matter fields (hopefully to be identified with the quarks and leptons of the standard model). In addition,  if we consider a model with open and closed  superstrings, or a model of heterotic superstrings   (i.e., closed strings in which only right-moving modes are supersymmetrized, while left-moving modes are not), we can directly include into the quantum spectrum also non-Abelian gauge fields,  with gauge group  $SO(32)$ or $E_8 \times E_8$ \cite{13,13a}. These superstring models,  when quantized in a flat space-time manifold, are only consistent for a critical number $D=10$ of dimensions. Remarkably, when $D-4=6$ spatial dimensions are appropriately compactified, the above gauge groups could possibly contain a realistic (low-energy) standard-model phenomenology. 

\subsection{Conformal invariance}

But the most impressive aspect of string theory, for a physicist used to work with the standard (classical or quantum) field theory, is probably the ability of fixing in a complete and unique way the equations of motion of all fields (bosonic and fermionic, massless and massive) present in the spectrum. This means, in other words, that if we accept superstrings  as realistic models of all existing fundamental  interactions, these models not only tell us that in Nature must exist, for instance, gravitational fields, Abelian and non-Abelian gauge fields, but also give us {\em the dynamical equations satisfied by these fields} -- and, at large distances,  these equations miraculously reduce to the Einstein equations for the gravitational field, and to the Maxwell and Yang-Mills equations for the gauge fields. 

This property of string theory probably represents the most revolutionary aspect with respect to theoretical models based on the notion of elementary  particle: the motion of a point-like test body, even  if quantized, does not impose in fact any restriction on the external fields in which the body is embedded and with which it interacts. 
Such background fields can satisfy arbitrarily prescribed equations of motion, usually chosen on the grounds of phenomenological indications: we can think, for instance, to the Maxwell equations, empirically constructed from the laws of Gauss, Lenz, Faraday and Ampere. It would be possible, in principle, to formulate different sets of equations still preserving the Lorentz covariance and other symmetry properties (such as the $U(1)$ gauge symmetry) typical of the electromagnetic interactions. Such different equations would be possibly discarded only for their disagreement with experimental results. In a string theory context, on the contrary, such alternative equations must be discarded {\em a priori}, as they would be inconsistent with the quantization of a charged string interacting with 
an external electromagnetic field. 

The above property of string-theory is grounded on the fact that the string action functional, representing the area of the two-dimensional world-sheet surface spanned by the string motion, is classically invariant under conformal transformations, (i.e., local deformations of the two-dimensional world-sheet metric $\ga_{ab}$, $a,b=0,1$). Thanks to this invariance we can always introduce the so-called ``conformal gauge" (where the intrinsic metric of the world sheet is the flat Minkowski metric, $\ga_{ab}=\eta_{ab}$), and we can always eliminate the ``longitudinal" string modes, leaving as a complete set of independent degrees of freedom only the modes describing oscillations transverse to the string. The conformal invariance plays thus a crucial role in the process of determining the correct quantum spectrum of physical string states. 

When we have a test string interacting with any one of the fields present in its spectrum (for instance the dilaton field, or the gravitational field, or a gauge field if the string is charged), we must then require, for consistency, that the conformal invariance (determining the string spectrum) be preserved by the given interaction, not only at the classical but also at the quantum level. 
This means, in other words, that the quantization of a string model including background interactions must avoid the presence of ``conformal anomalies", i.e. of quantum violations of the conformal invariance already present at the classical level. It follows that  the only background-field configurations admissible in a string theory context are those satisfying the conditions of conformal invariance. Such conditions are represented by a set of differential equations corresponding, in every respect, to  the equations of motion of the field we are considering.

Unfortunately, however, such equations can hardly be derived in closed and exact form for any given model of interacting  string. In practice, we have to adopt a perturbative approach: the action of the string interacting with the background fields (also called ``sigma model" action) is quantized by expanding in loops the quantum corrections (as in the standard field-theory context, but with the difference that we are dealing with a two-dimensional conformal field theory). The absence of conformal anomalies is then imposed at any order of the loop expansion, determining the corresponding differential conditions. As a  result, the exact background  field equations are approximated by an infinite series of equations, containing higher and higher derivative terms as we consider loops of higher and higher order. 

To lowest order we then recover the second-order differential equations already well known for the classical fields (the Maxwell, Einstein, Yang-Mills equations, as well as the Dirac equations for the fermion fields). To higher orders there are quantum corrections to these equations, in the form of higher derivatives of the fields, 
appearing as an expansion in powers of the parameter  $\ls^2$.  
The corrections to the equations of motion of order $\ls^2$ and higher are a typical effect of the theory due to the finite extension of strings: indeed, such corrections disappear in the point-particle limit 
$\ls \ra 0$, while they become important in the strong field/higher curvature limit in which the  length scale of a typical process (for instance, the space-time curvature scale) becomes comparable with the string scale  $\ls$. In the context of pre-big bang cosmology such corrections can play an important role in the transition to the phase of standard decelerated evolution, as we shall see in Sect. \ref{sec4}. 

\subsection{Gravitational and cosmological applications}

Let us now come back to one of the main application of string theory in a gravitational context, concerning the problem of spacetime singularities. Unfortunately, technical difficulties have prevented, so far, to reach clear-cut conclusions on the fate of the cosmological singularity (unlike other kinds of singularities). By contrast, considerable progress was made on quantum black hole physics, in particular a microscopic (statistical mechanics) understanding of their Bekenstein-Hawking  entropy was achieved for a particular class of black holes \cite{blackhole}. Also, the study of (gedanken) superplanckian string collisions \cite{stringcollision} has led to the conclusion that, in string theory, black holes have a minimal size corresponding to a maximal Hawking temperature  of the order of string theory's maximal temperature \cite{temp, temp1}.  
An obvious interpretation of this result stems from the simple   observation that fundamental quantum strings, unlike their classical analogs, have a minimal (or better optimal) size, $\ls$. This physical argument strongly suggests a cosmological analogue of the above conclusion. Because of their finite size quantum strings cannot occupy a vanishing volume, suggesting an upper limit to the energy density and the  spacetime curvature. If so, string theory should be able to avoid -- or reinterpret -- the big bang singularity predicted by classical general relativity.

Other properties of quantum strings that could play an important role in a cosmological context are:

$(a)$ {\em Some new symmetries characterizing string theory as opposed to (classical or quantum) field theory.} 

$(b)$  {\em The necessity of new dimensions of space.} 

$(c)$ {\em The existence of new light fields, at least within perturbation theory.}\\
Let us comment briefly on each one of them in turn.

1. An example of a genuine stringy symmetry is the so-called target-space duality (or  $T$-duality) \cite{13b}. It stems from quantum mechanics since it is due to the possibility of interchanging the ``winding" modes of a closed string (winding number being classically an integer, counting the number of times a string is wrapped around a compact spatial dimension) with its momentum modes (which in a compact direction are discrete, thanks to quantum mechanics). A cosmological variant of $T$-duality is heavily used to motivate an example of pre-big bang cosmology, as we shall see in Sect. \ref{sec4}.

2. If we wish our string theory to allow for smooth, Minkowski-like solutions, we have to allow for more than three dimensions of space. At weak coupling six extra dimensions are needed, while at strong coupling the dilaton  field (that controls the coupling) can be reinterpreted as a seventh extra dimension of space (for a total of $D=11$ spacetime dimensions). Although one would like to have all the extra dimensions to be small and static during the recent history of our  Universe, it is all but inconceivable that extra dimensions may have played some role in the very early (e.g. pre-big bang) Universe. An example of making good use of them in order to produce a realistic spectrum of density perturbations will be given in Sect. \ref{sec4}

3. String theory does $\it{not}$ automatically give  general relativity at large scales. Besides the presence of the extra dimensions, string theory also contains, at least  in perturbation theory (i.e. at weak coupling), many massless scalar particles in the spectrum that may   induce unobserved long range forces of gravitational (or even stronger) strength. 
These particles, and their interactions, are described by the low-energy string effective action which, as already stressed,  controls the 
predictive power of the theory at the experimentally accessible energy scales (at least if the string and Planck scale are of comparable magnitude). 

One of these particles is the above-mentioned dilaton. Its massless nature reflects the fact that, in perturbation theory, the so-called string coupling $g_s$ is a free parameter (see below for a short explanation of the connection between the string coupling and the dilaton). Other (perturbatively) massless particles are those controlling the sizes and shapes of the extra dimensions. All these unwanted massless particles are generically called ``moduli", and the problem associated with them is called the ``moduli stabilization problem". It is generally felt that the mechanism of spontaneous supersymmetry breaking, when correctly implemented, will produce a non-perturbative mass for the moduli and thus stabilize their value. If, however, as suggested by $T$-duality, the universe started at very weak coupling, the moduli may have been all but static in the early Universe, and may have affected its evolution in an essential way. Consequences of such an evolution should be observable even today, as we shall discuss in Sect. \ref{sec4}.

\subsection{Topological expansion}

It seems finally appropriate, also in view of the subsequent cosmological applications, to explain the close connection between the string coupling parameter $g_s$ and the dilaton. Let us recall, to this purpose, that the higher-derivative expansion in powers of $\ls^2$, introduced previously, is not the only perturbative approximation conceivable in a string theory context. Another useful expansion of the equations of motion of the background fields concerns the topology of the two-dimensional world-sheet spanned by the string: it is the so-called ``higher genus" expansion. 

\begin{figure}
\centering
\includegraphics[height=3cm]{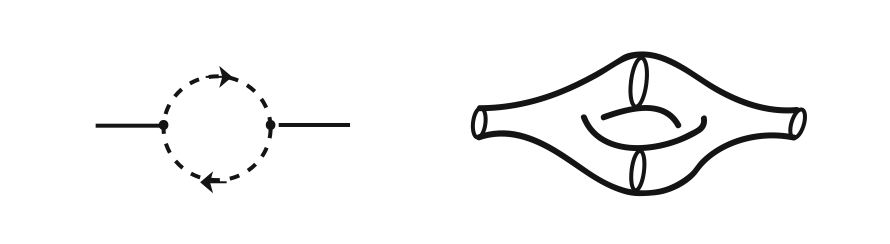}
\caption{One-loop graph for a point particle (left) and a closed string (right). In the left picture the world-line of a physical point particles (solid curve) splits into a ``world-loop" (dashed curve) representing a virtual particle-antiparticle pair generated by the quantum fluctuations of the vacuum. In the right picture, where the world-lines are replaced by cylindrical ``world-sheet" surfaces,  the same process is illustrated for the case of a closed string.}
\label{fig1}       
\end{figure}

Consider for instance a closed string, whose propagation in the target-space manifold describes a cylindrical world-sheet surface. If the string splits into two strings, which subsequently recombine to form again the initial string (with a process analogous to the one described by a one-loop Feynman graph in quantum field theory), the world-sheet will acquire the topology of a torus (see Fig. \ref{fig1}). A process with $n$ loops will correspond, in general,  to a two-dimensional Riemann surface of genus $n$, i.e. to a manifold with $n$ ``handles". The interactions among strings can then be described, within a perturbative approach, by a partition function which can be expanded in world-sheet configurations of higher and higher genus (i.e., higher and higher number of handles).

For a two-dimensional, closed and orientable manifold, on the other hand, the genus $n$ is determined by the so-called Euler characteristic $\chi= 2-2n$. This quantity is a  topological invariant which (by virtue of the Gauss-Bonnet theorem) can be expressed as an integral over the intrinsic scalar curvature of the world-sheet surface. The dilaton, by definition, appears in the action $S$ as multiplying such a scalar curvature; hence, for a constant dilaton,  
$S \sim  (\phi/2)\chi $. The partition function, on the other hand, contains $\exp(-S)$. An expansion of the partition function in a series of higher-genus world sheets thus becomes an expansion in powers  {\em of the exponential of the dilaton field}, $ \exp  \phi $. But, by definition, the loop approximation is also an expansion in powers of the coupling constant  $g_s^2$: this immediately gives the (perturbative) relation between a constant dilaton and the string coupling parameter, namely $g_s^2= \exp \phi$. Even if $\phi$ is not a constant, $\exp \phi$ still plays the role of a {\em local} effective coupling.

\section{String theory and inflationary cosmology}
\label{sec3}

As discussed in the previous section, superstring theory seems to provide complete and consistent models for all the fundamental components of matter and all interactions, valid at all energy scales: thus, it can be appropriately applied to describe the evolution of the primordial Universe in the regime in which all interactions were unified and possible quantized (gravity included). We can then investigate, in this context, the possible (spontaneous?) occurrence of a phase of accelerated inflationary evolution, able to provide the correct initial conditions for the subsequent, standard regime. 
We can ask, in particular, whether string theory has a natural candidate for the role  of the inflaton, and for driving a successful phase of slow-roll inflation, without inventing {\em ad hoc} a new field characterized by  the required properties. 

The answer to the above questions would seem to depend on the particular type of superstring model chosen to describe the primordial Universe: there are indeed five different models (see e.g. \cite{13,13a}), related by duality transformations, which probably describe different physical regimes of the same underlying theory (the so-called M-theory), and which are characterized by a significantly different field content. All these models, however, contain a fundamental scalar field, the dilaton, which is coupled to gravity always in the same way, and which leads to an effective model of scalar-tensor gravitational interactions. This scalar field is expected to  acquire a non-perturbative  potential in the strong coupling regime $g_s^2 \sim 1$:  any superstring model thus automatically contains, in principle, all the required ingredients for the formulation of an inflationary scenario based on the dynamics of a self-interacting scalar field. 

Unfortunately,  a dilaton potential typical of string theory  (without contrived and/or {\em ad hoc} modifications) seems to be unsuitable for implementing a successful model of slow-roll inflation, as shown by various studies at the beginning of the Nineties \cite{14,14a}. We are thus left with two possible alternatives:  

$(i)$ {\em look for an inflationary scenario based on the dilaton,  different from the conventional slow-roll inflation}; or 

$(ii)$ {\em look for another mechanism, independent of the dilaton, and  able to implement a phase of slow-roll inflation with some other typical ingredient of string theory.} 

The first approach, proposed more than twenty years ago, has lead to the pre-big bang scenario that we shall present in the following section. The second approach is more recent, and has lead to scenarios where the inflaton field corresponds to the distance between two three-dimensional membranes (called $3$-branes), propagating through a higher-dimensional space-time manifold. 

This second possibility has been implemented, in particular, considering the interaction between a brane and an anti-brane, which attract each other because of their opposite ``gravitational charges" \cite{15} (see \cite{17} for a detailed introduction of the so-called ``brane-antibrane" inflationary mechanism). But there are also examples of brane-cosmology in which the two $3$-branes are ``domain walls" representing the space-time boundaries: we are referring to the so-called ``ekpyrotic" \cite{18,18a} or ``new ekpyrotic" \cite{burt} scenario, where the brane interaction (and their eventual collision) is associated to the shrinking of an extra spatial dimension orthogonal to the branes. One should also mention the so-called models of ``multi-brane" inflation \cite{Axel}, where several M-theory branes combine their steep interactions to produce an acceptable (i.e., flat enough) inflaton potential as in models of ``assisted" inflation \cite{Liddle}.

It is important to recall, in this context, that all models of brane-antibrane inflation --   either based on a non-trivial topology \cite{15} or on a non-trivial geometry of the transverse dimensions \cite{20} -- are consistent provided that shape and volume of the extra (non-inflationary) spatial dimensions are stabilized, using some appropriate mechanism \cite{21} which does not affect the relative motion of the two branes. The stabilization mechanism, in turn, seems to require the presence of antisymmetric tensor fields,  whose ``fluxes" are associated to branes ``wrapping" around around the compact dimensions and ``warping" the compactified geometry \cite{22}. This enforces the conclusion that the above realizations of slow-roll inflation are only possible when the Universe is in a phase of ``brane-domination". 

On the other hand, in a string theory context, the tension of a $D$-brane (i.e. the mass per unit of spatial (hyper)volume) is proportional to the inverse of the string coupling parameter $g_s$: it follows that such branes  become light, and can  be copiously produced, only in the strong coupling limit $g_s \gg1$. A phase of brane-antibrane inflation seems thus to be naturally (and unavoidably) associated to a phase of strong coupling. 

We are then lead to the question: how to ``prepare" the strong-coupling (and, possibly high-curvature) regime where the mechanism of brane  inflation can be implemented? In other words, what happens ``before"? What about the cosmological evolution before the beginning of such an inflationary regime? We know, in fact, that slow-roll inflation cannot be past-eternal: going back in time, at fixed value of the gravitational coupling, a phase of slow-roll inflation necessarily leads to a spacetime (big bang) singularity in a finite amount of proper time, and the model of cosmological evolution remains incomplete. Is it  possible to arrive at the brane-dominated regime, where the Universe possibly undergoes a phase of slow-roll inflation, without starting from an initial singularity?  

The pre-big bang scenario,  which,  as we shall see  in the following section, was originally suggested as an alternative inflationary mechanism, may also provide answers to these questions.

\section{The pre-big bang scenario}
\label{sec4}

As we mentioned in the introduction a basic assumption underlying (this version of) string cosmology is that string theory, thanks to the fundamental length scale ($\ls$) it contains, avoids the initial singularity of Classical General Relativity thus allowing for a phase preceding the high temperature, high density and high curvature state in which the Universe must have been some 15 billion years ago. But what did the Universe look like in that pre-bangian phase?

An appealing possibility is that the Universe did not have a real beginning and that, instead, by going towards the infinite past, it approached the simplest possible state compatible with the field equations of superstring theory: a flat 10-dimensional space with a constant and very negative dilaton field. This state is nothing but the trivial perturbative vacuum of string theory with vanishing curvature and  coupling ($g_s^2 \sim e^{\phi} \rightarrow 0$).

Such a trivial initial state looks very special and thus highly fine-tuned. However, this is not the case if we realize that the assumption refers to the {\it asymptotic} past. In other words, we may assume that the initial state is a {\it generic} solution to the metric plus dilaton equations of motion (in the technical sense that it contains the correct number of arbitrary functions) approaching asymptotically the trivial vacuum.
This has been termed the assumption of  `` Asymptotic Past Triviality" in \cite{31}.

As we shall now discuss an initial state of this type, flat, cold, empty and decoupled, is just what one gets under the hypothesis that the Universe evolves in a ``self-dual" way with respect to a particular symmetry of the low-energy string effective action \cite{11,12}.

\subsection{Scale-factor duality}

The symmetry we have in mind is the so-called scale-factor duality \cite{12,23}, which generalizes to time-dependent backgrounds the $T$-duality symmetry mentioned in Sect. \ref{sec2}, typical of closed strings in static manifolds with compact spatial dimensions  \cite{13b}. According to the $T$-duality symmetry the energy spectrum of a string, quantized in the presence of compact dimensions of radius $R$, turns out to be invariant with respect to the transformation $R \ra \ls^2/R$, and the simultaneous exchange of the number of times the string is wound around the compact dimensions with the (discrete) level-number of the momentum operator associated with the string motion along such compact directions. This transformation must be accompanied by a transformation of the (constant) dilaton such as to keep the effective coupling in the uncompactified dimensions unchanged.

In the context of a homogenous, isotropic, spatially flat background, described by the invariant space-time interval  
\beq
ds^2= c^2dt^2- a^2(t)\, dx_idx^i,
\label{3}
\eeq
the radius $R$ is replaced by the scale factor $a(t)$, and the invariance under the transformation $a \ra a^{-1}$ is still valid provided the dilaton field $\phi$ is also simultaneously transformed, according to the 
 (scale-factor duality) transformation
\beq
a \ra \ti a = a^{-1}, ~~~~~~~~~~
\phi \ra \ti \phi = \phi- 2 d \ln a, ~~~~~~~~\fb \equiv \phi - d \ln a \ra \fb
\label{4}
\eeq
where $d$ is the total number of spatial dimensions and $\fb$ is the so-called shifted dilaton (left invariant under scale-factor duality). There are, however, two important differences between $T$-duality and 
scale-factor duality symmetry: $i)$ the above transformations represent a symmetry of the (tree-level) equations of motion of the background fields, and not necessarily of the quantum string spectrum; $ii)$ there is no need of compact spatial dimensions. 

The invariance of the string cosmology equations under the transformations (\ref{4}) can be extended to include the presence of matter sources, as we shall see below. Furthermore such an invariance is only a particular case of a more general class of symmetries (involving also the $B_{\mu \nu}$ field) associated twith a global $O(d,d)$ group, for backgrounds with $d$  Abelian isometries. This extended symmetry of the cosmological equations of string theory holds both  without \cite{24} and with \cite{25} matter sources. For the purpose of this paper, however, it will be enough to limit ourselves to the transformations of Eq. (\ref{4}), and to note that new classes of solutions can be obtained simply by transforming known solutions. These new solutions have kinematic properties representing  the ``dual counterpart" of the initial ones. Such a a duality symmetry should also leave an imprint on the properties of the cosmological perturbations 
\cite{17a}.

For an explicit illustration of this duality symmetry, and of its applications, we need the cosmological equations obtained  from the low-energy string effective action, describing the dynamics of the graviton and dilaton field to the lowest non-perturbative order. Limiting for simplicity our discussion to the case of $d=3$ non trivial dimensions (with the other six being compct and static) such equations, for the metric (\ref{3}) and for a homogeneous dilaton field $\phi=\phi(t)$, take the form (from now on we shall adopt units in which $\hbar=c=1$):
\bea
&&
\fpu^2-6 H \fpu+ 6 H^2= {2 \ls^2} e^\phi \r,
\nonumber \\ &&
\dot H- H \fpu+ 3H^2= {\ls^2} e^\phi p,
\nonumber \\ &&
2 \fpp-\fpu^2 + 6 H \fpu - 6 \dot H -12 H^2 =0
\label{5}
\eea
(see e.g. \cite{11} or \cite{26}). 
We are working here in the so-called string frame (where the motion of a test string is geodesics), and the dot denotes differentiation with respect to the cosmic time $t$ (we also recall that $H= \dot a/a$). We have also taken into account the possibility of matter sources, in the form of a perfect fluid with stress tensor $T_0\,^0=\r$ and $T_i\,^j= -p \da_i^j$. Note that, in four space-time dimensions, the combination $(2\ls^2)\exp \phi$ plays the role of the effective gravitational coupling constant $16 \pi G= 2 \lp^2$. Thus, the dilaton controls the relative string-to-Planck length ratio, to this approximation, as
\beq
\left(\lp\over \ls\right)^2 =\left(\Ms\over \Mp\right)^2= e^\phi= g_s^2.
\label{6}
\eeq

The above set of cosmological equations is clearly invariant (as in the case of the Einstein equations) under the time-reflection transformation $t \ra -t$. In addition, however, the equations are invariant under the duality transformations (\ref{4}), possibly accompanied by the following transformations of the matter sources \cite{12}:
\beq
\r \ra \ti \r = \r \,a^6, ~~~~~~~~~~
p \ra \ti p = - p\,  a^6.
\label{7}
\eeq
Thanks to this symmetry we can always associate to any decelerated solution, typical of the standard cosmological model, a ``dual partner" describing accelerated (i.e. inflationary) evolution. We stress that this is impossible in the context of the cosmological Einstein equations, where there is no dilaton, and no duality symmetry. 

Consider, for instance, a radiation-dominated Universe, characterized by the equation of state $p= \r/3$, and by a constant dilaton, $\fpu=0$. In  this case the set of Eqs. (\ref{5}) is identically satisfied by the particular exact solution
\beq
a=\left(t\over t_0\right)^{1/2}, ~~~~~ \phi= {\rm const}, ~~~~~ 
\r= {\r_0\over a^4}, ~~~~~ p= {\r\over 3}, ~~~~~ t>0,
\label{8}
\eeq
where $t_0$ and $\r_0$ are positive integration constants. This is also the exact solution of the conventional Einstein equations which, for $t>0$, properly describes the radiation-dominated phase of the standard cosmological scenario \cite{1}. This solution describes a phase of decelerated expansion, decreasing curvature and constant dilaton for 
$0 < t \leq \infty$, 
\beq
\dot a >0, ~~~~~ \ddot a <0, ~~~~~ \dot H <0, ~~~~~ \fpu=0,
\label{9}
\eeq
namely a Universe evolving from the past big bang singularity (approached as $t \ra 0$ from positive values) towards  a state asymptotically flat and empty, approached as  $t \ra +\infty$. 

By applying a combined scale-factor duality and time-reversal transformation on the solution (\ref{8}) we can now obtain a new exact
solution of the low-energy string cosmology equations (\ref{5}) which has the form
\beq
\ti a=\left(-{t\over t_0}\right)^{-1/2}, ~~~~ \ti \phi=-3 \ln \left(-{t\over t_0}\right) , ~~~~~ 
\ti \r= {\r_0\over \ti a^2}, ~~~~~ \ti p= -{\ti \r\over 3}, ~~~~~ t<0.
\label{10}
\eeq
This new solution is defined on the negative time semi-axis, $- \infty \leq t <0$, and describes an (inflationary) phase of accelerated expansion, growing curvature and growing dilaton, 
\beq
\dot a >0, ~~~~~ \ddot a >0, ~~~~~ \dot H >0, ~~~~~ \fpu>0.
\label{11}
\eeq
As anticipated at the beginning of this section the initial cosmological state is asymptotically flat,  $H \ra 0$, with esponentially suppressed string coupling, $g_s^2 = \exp \phi \ra 0$, as $t \ra -\infty$ (the Universe approaches, asymptotically, the so-called ``string perturbative vacuum"). Starting from this configuration, the evolution leads  towards a future singularity approached as $t \ra 0$ from negative values. 

\subsection{Examples of smooth pre-big bang backgrounds}

The two exact solutions (\ref{8}) and (\ref{10}) cannot be directly connected to one another, being physically separated by a curvature singularity at $t=0$. Assuming however that the singularity may disappear in a more realistic model, after including other fields and/or higher-order corrections, we obtain from the duality invariance of the equations a clear suggestion about a possible ``temporal completion" 
of the standard cosmological phase, based on a  ``self-duality principle". 

\begin{figure}
\centering
\includegraphics[height=5cm]{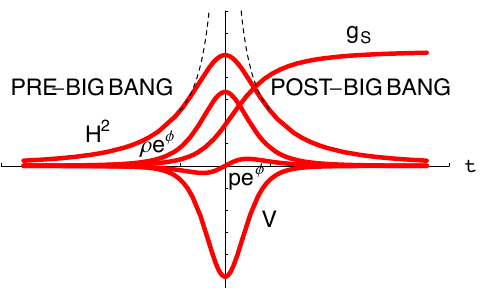}
\caption{Example of smooth transition between a phase of pre-big bang inflation and the standard radiation-dominated evolution. The figure gives a qualitative illustration of the evolution in time (from $-\infty$ to $+\infty$) of the string coupling parameter $g_s= \exp(\phi)$, of the spacetime curvature scale $H^2= \dot a^2/a^2$, of the effective energy density $\r \exp(\phi)$ and pressure $p \exp(\phi)$ of the gravitational sources, and of the effective dilaton potential $V$. Note that the pre-big bang phase is characterized by growing curvature and negative pressure, the post-big bang phase by decreasing curvature and positive pressure.}
\label{fig2}       
\end{figure}

Indeed, if the scale factor would satisfy (even approximately) the condition  $a(t)= a^{-1}(-t)$, then the standard cosmological phase would be preceded and completed by a dual phase which is automatically of the inflationary type (see for instance Eq. (\ref{10})). The space-time curvature scale (growing during inflation, decreasing during the decelerated evolution) would have a specular behavior with respect to a critical ``bouncing epoch" centered around $t=0$ (namely would be reflection-symmetric in the $\{t, H\}$ plane with respect to the $t=0$ axis, marking the transition from the inflationary to the standard regime; see e.g. Fig. \ref{fig2}). Such a bouncing transition, occurring at high but finite curvature, would replace the big bang singularity of the low-energy equations: thus, one is naturally lead to call ``pre-big bang" the initial phase (at growing curvature and string coupling), with respect to the subsequent ``post-big bang" phase describing standard decelerated evolution. 

Regular, self-dual solutions  cannot be obtained \cite{26a} in the context of the low-energy string-cosmology equations (\ref{5}). Such equations, however, only represent a zeroth-order approximation to the exact field equations expected in a string theory context, as already stressed in Sect. \ref{sec2}. On the other hand, if the curvature  and the string coupling keep growing without any damping during the phase described by the low-energy pre-big bang solutions, then the Universe is necessarily driven to configurations where the curvature and the coupling are large in string units ($\ls^2  H^2 \sim 1$, $g_s^2 \sim 1$): for those configurations both higher-derivative and higher-loop corrections are to be added (possibly to all orders) to the tree-level equations. All studies performed up to now \cite{27,27a,27b} (see also \cite{26}) have shown that such corrections can indeed contribute to damp the accelerated evolution of the background, and trigger the transition to the post-big bang phase. 

Without introducing complicated computations we will present here a phenomenological example of self-dual solutions based on a special class of dilaton potential, depending on $\phi$ through the previously defined shifted dilaton ($\fb= \phi - 3 \ln a$ for $d=3$). Potentials of that type are motivated by the invariance of $\fb$ \cite{12,24,25}; in addition, they could be generated by dilaton loop corrections in manifold with compact spatial sections, and can be obtained in the homogeneous limit of appropriate general-covariant (but non-local) actions \cite{28}. With such potential $V= V(\fb)$ the string cosmology equations can be rewritten in terms of $a$, $\fb$, $\rb= \r a^3$ and $\pb= p a^3$ as follows \cite{24,25}:
\bea
&&
\fbp^2 -3 H^2-V(\fb)= {2 \ls^2} e^{\fb} \,\rb,
\nonumber \\ &&
\dot H- H \fbp =  {\ls^2} e^{\fb} \,\pb,
\nonumber \\ &&
2 \fbpp-\fbp^2 - 3 H^2 +V(\fb)- {\pa V \over \pa \fb}=0.
\label{12}
\eea

These equations are still invariant under the duality transformations (\ref{4}) and (\ref{7}) but, differently from Eq. (\ref{5}), they admit regular and self-dual solutions. We can also obtain exact analytical integrations for appropriate forms of the potential $V(\fb)$, and for equations of state such that $p/\r$ can be written as integrable function of a suitable time parameter \cite{11}. 

Let us consider, as a simple example, the exponential potential $V= -V_0 \exp (2 \fb)$ (with $V_0>0$), to be regarded here only as an effective, low-energy description of the quantum-loop backreaction, possibly computable at higher orders. Let us use, in addition, an equation of state (motivated by analytical simulations concerning the equation of state of a string gas in backgrounds with rolling horizons \cite{29}) evolving between the asymptotic values $p= - \r/3$ at $t \ra -\infty$ and $p=  \r/3$ at $t \ra +\infty$, so as to match the low-energy pre-and post-big bang solutions (\ref{10}) and (\ref{8}), respectively. The plot of the corresponding solution (see \cite{11} for the exact analytic form) is illustrated in Fig. \ref{fig2}.

The solution smoothly interpolates between the string perturbative vacuum at $t \ra -\infty$ and the  standard, radiation-dominated phase at constant dilaton (described by Eq. (\ref{8})) at $ t \ra + \infty$, after a pre-big bang phase of growing curvature and growing dilaton described by Eq. (\ref{10}) (the growth of the coupling can be in principle accompanied also by a large amount of entropy production \cite{20a}).
The dashed curves of Fig. \ref{fig2} represent the unbounded growth of $H$  in the low-energy solutions (\ref{8}), (\ref{10}), obtained in the absence of the dilaton potential. The plot of $V$ in Fig. \ref{fig2} shows that the potential contribution dominates the background evolution around the transition epoch $|t| \ra 0$, and rapidly disappears at large values of $|t|$. 

An even simpler (almost trivial) example is obtained, with the same potential, in the case $p=0$. The system of equations (\ref{12}) is then satisfied by a flat geometry ($H=0= \dot H$) and by a ``bell-like" time evolution of the dilaton, sustained by a constant energy density: we find indeed the very simple solution \cite{42}
\beq
a =a_0, ~~~~~ \r=\r_0, ~~~~~ p=0, ~~~~~ e^\phi= {e^{\phi_0}\over 1 +(t/t_0)^2} ,
\label{13}
\eeq
where the integration constants $a_0$, $\phi_0$, $\r_0$, $t_0$ are related by $\r_0 \exp \phi_0= V_0 \exp(2 \phi_0)=4/t_0^2$. For $t \ra -\infty$ one has a globally flat initial Universe, filled with a constant and uniform distribution of ``dust" dark matter.

\begin{figure}
\centering
\includegraphics[height=5cm]{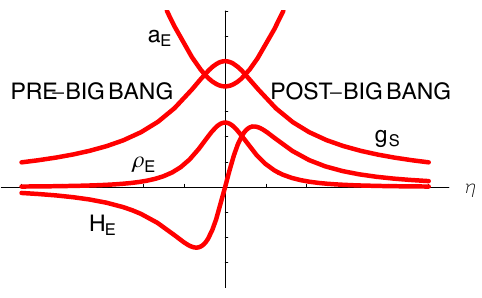}
\caption{Example of pre-big bang evolution represented in the E-frame, where the scale factor $a_E$ is shrinking and the Hubble parameter $H_E$ is negative, unlike in  the string-frame representation of Fig. \ref{fig2}. The figure also illustrates the evolution (with respect to the  conformal time parameter $\eta$) of the E-frame energy density $\r_E$ and of the string coupling parameter $g_s =\exp(\phi)$. 
The plots are obtained from Eq. (\ref{14}) with $a_0=0.8$, $\phi_0=0$, $\r_0=1$, $\eta_0=1$.}
\label{fig3}       
\end{figure}

The solution is less trivial, however, when transformed to the Einstein-frame (E-frame) where the graviton and the dilaton are canonically normalized, and where the metric and the energy density are no  longer constant. Using the conformal time parameter $\eta$, the E-frame version of the solution (\ref{13}) takes the form \cite{28} (see Fig. \ref{fig3})
\beq
~~~a_E= a_0 e^{-{\phi_0\over 2}} \left(1+{\eta^2\over \eta_0^2}\right)^{-{1\over 2}},
~~~e^\phi=e^{\phi_0} \left(1+{\eta^2\over \eta_0^2}\right)^{-1},
~~~\r_E= \r_0 a_0^3 \left(1+{\eta^2\over \eta_0^2}\right)^{-{3\over 2}}.
\label{14}
\eeq
In Fig. \ref{fig3} we have plotted the exponential of the dilaton field,  the E-frame scale factor $a_E$, the energy density $\r_E$, and the Hubble parameter $H_E=a'_E/a_E^2$ (where $a'= d a/d \eta$), obtained from Eq. (\ref{14}). The background smoothly evolves from growing to decreasing curvature (and dilaton), but the pre-big bang regime corresponds, in the E-frame, to a phase of {\em accelerated contraction} ($H_E<0$, $\dot H_E <0$), with a final bounce of the scale factor to the phase of post-big bang (decelerated) expansion. 

\begin{figure}
\centering
\includegraphics[height=6cm]{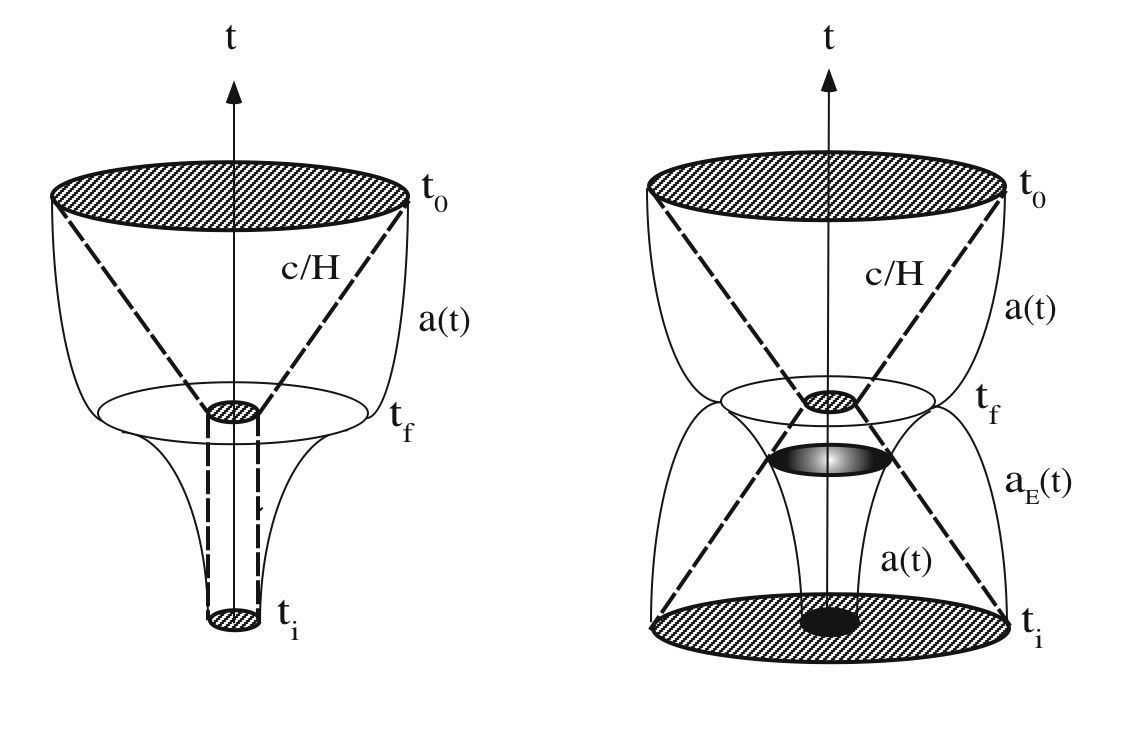}
\caption{Comparison between the time evolution of the Hubble horizon  $c/H$ (dashed lines) and of the scale factor $a(t)$ (solid curves) in the conventional inflationary scenario (left) and in models of pre-big bang inflation (right).  For the pre-big bang phase we have plotted the evolution of both the expanding string-frame scale factor $a(t)$ and the contracting E-frame scale factor $a_E(t)$. The vertical axis is the time axis, and the shaded areas represent causally connected spatial sections of Hubble size $c/H$ at various epochs. The evolution from the end of inflation, $t_f$, to the present epoch, $t_0$, is the same in both cases. However, during inflation (i.e. from $t_i$ to $t_f$) the Hubble horizon is constant (or slightly increasing) in conventional models (left), while it is shrinking in pre-big bang models (right).  As a consequence, the size of the initial inflationary patch may be very large (in string or Planck units) for a phase of pre-big bang inflation, but {\em not} larger than the horizon itself, as illustrated in the figure.}
\label{fig4}       
\end{figure}

As clearly illustrated by the above examples, according to the pre-big bang scenario inflation starts  in a perturbative regime of very small coupling ($g_s \ra 0$), very small curvature ($H \ra 0$), and very large space-time horizons $cH^{-1}$. The trans-Planckian problem (mentioned in Sect. \ref{sec1}) is therefore neatly evaded, and the initial cosmological evolution can be consistently described by a classical, low-energy effective action. This represents a complete overturning of the traditional picture -- typical of standard models of inflation \cite{30} -- where the initial evolution takes place at ultra-high density and curvature scales, 
small horizon scale, strong coupling, in a marked quantum regime (see Fig. \ref{fig4}). 

As we have already mentioned, it is possible to characterize these new initial conditions in terms of a principle of ``Asymptotic Past Triviality" \cite{31} (APT) according to which the Universe becomes empty, flat, and interaction-free as one goes towards the asymptotic past,  $t \rightarrow- \infty$. It was argued in \cite{31} that, thanks to gravitational instability,  ``generic"  initial conditions satisfying APT unavoidably lead to the formation of black holes of many different masses/sizes. It was also shown that the spacetime inside  such black holes  precisely mimics the pre-bounce phase of a pre-big bang cosmology. If the singularity inside the black hole horizon is avoided, a sufficiently large black hole can then give rise to a successful pre-big bang cosmology.

In conclusion, an epoch of pre-big bang inflation is able to solve the kinematical problems of the standard scenario starting from different initial conditions which are not necessarily unnatural  or unlikely \cite{31,32} (see also \cite{33} for a detailed comparison of the pre-big bang versus post-big bang inflationary kinematics). A possible exception concerns the presence of primordial ``shear", which is not automatically inflated away during the phase of pre-big bang evolution: the isotropization of the three-dimensional spatial sections might require some specific post-big bang mechanism (see e.g. the discussion of \cite{33a}), unlike in the standard inflationary scenario where the dilution of shear is automatic. 

\subsection{Phenomenological consequences}

Quantum effects, in the pre-big bang scenario, can become important towards the end of the inflationary regime. We can say, in particular, that the monotonic growth of the curvature and of the string coupling automatically ``prepares" the onset of a typically ``stringy" epoch at strong coupling. This epoch could be characterized  by the production of a gas of heavy objects (such as winding strings \cite{34,34a} or mini-black holes \cite{53a}) as well as light, higher-dimensional branes \cite{34b} (see \cite{Wat} for a review of the so-called phase of ``string gas" cosmology). In such a context the interaction (and/or the eventual collision) of two branes can drive a phase of slow-roll inflation \cite{15}, as discussed in Sect. \ref{sec3}. 

At this point of the cosmological evolution there are two possible alternatives. 

$(i)$ The phase of string/brane dominated inflation is long enough to dilute all effects of the preceding phase of dilaton inflation, and to give rise to an epoch of slow-roll inflation able to prepare the  subsequent evolution, according to the conventional inflationary picture. 

$(ii)$  The back-reaction of the quantum fluctuations, amplified by the phase of pre-big bang inflation, induces a bounce as soon as the Universe reaches the strong coupling regime; almost immediately, there is the beginning of the phase of decelerated evolution dominated by the (rapidly thermalized) radiation produced by the bouncing transition. 

These two possibilities have different impact on the phenomenology of the cosmic background of gravitational radiation and on the anisotropy of the CMB radiation, as we shall now discuss. 

In the first case (provided it can be successfully realized)  we would recover the same phenomenology of conventional models of slow-roll inflation, but with the important difference that the inflationary initial conditions are now ``explained" by the preceding pre-big bang evolution, driving the Universe from the string perturbative vacuum to the appropriate configuration for starting inflation. The slow-roll singularity is eliminated and the model becomes geodesically complete, being extended over the whole cosmic-time axis. We can still talk of  ``birth of the Universe from the string perturbative vacuum", as pointed out (in a quantum cosmology context) in \cite{35,35a,35b}.

Here we shall concentrate our discussion on the second possibility, in which there is only one inflationary phase of unconventional (pre-big bang) type. In that case the effects of the higher-curvature/quantum gravity phase are not ``washed out" by a subsequent inflationary epoch: the phenomenological predictions are possibly less robust, as may depend on (still unknown) details of string/Planck scale physics. On the other hand, the observable consequences of this scenario should bear the direct imprint of the pre-big bang epoch, thus opening a window on time scales and energy scales not accessible to conventional model of inflation. 

In particular, because of the peculiar kinematic properties of pre-big bang inflation, the quantum fluctuations of the metric tensor tend to be amplified with an unconventional spectrum which is ``blue'', i.e. growing with frequency (as pointed out long ago in \cite{36,36a}), instead of being flat, or decreasing, as in the conventional inflationary scenario. A growing spectrum, on the other hand, simultaneously represents an advantage and a difficulty of pre-big bang models with respect to conventional models of inflation. 

\subsubsection{\it{Tensor metric perturbations}}

\begin{figure}
\centering
\includegraphics[height=5cm]{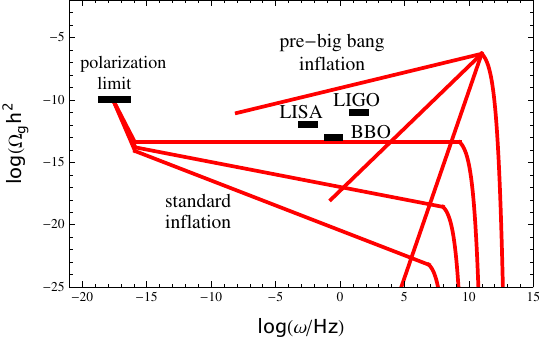}
\caption{The figure shows, on a logarithmic scale, the typical spectral energy density $\Om_g h^2$ of a cosmic background of primordial gravitational waves produced  in the context of $(a)$ minimal models of pre-big bang inflation (growing, or ``blue'', spectra) and $(b)$ standard models of slow-roll inflation (flat and decreasing, or ``red'', spectra). See \cite{26} for technical details on the properties of these spectra. The figure also shows the maximal background intensity allowed by present data on the CMB polarization, and 
the planned sensitivities expected to be reached in the near future by Earth-based gravitational detector such as Advanced LIGO and by space interferometers such as LISA and BBO.}
\label{fig5}       
\end{figure}

The (phenomenological) advantage concerns the amplification of the (transverse, traceless) tensor part of the metric fluctuations, and the corresponding formation of a cosmic background of relic gravitational radiation. In models of pre-big bang inflation, in fact, the spectral energy of such background -- $\Om_g h^2$, where $\Om_g$ is the energy density in critical units, and $h$ the present value of the Hubble parameter in units of $100$ km ${\rm s}^{-1} {\rm Mpc}^{-1}$ -- is expected to growth with frequency \cite{36,36a,36b}, with a peak value \cite{36c} of about $\Om_g h^2\sim 10^{-6}$, reached around the cut-off frequency of the spectrum ($\om \sim 100$ GHz, roughly corresponding to the position of the peak of the electromagnetic CMB radiation). This behaviour is in sharp contrast with the gravitational background produced in models of slow-roll inflation, characterized by a peak value which -- according to the most recent results on the polarization of the CMB radiation \cite{Bicep} -- is at most of order $\Om_gh^2 \sim 10^{-10}$
at the frequency scale of the present Hubble radius ($\om \sim 10^{-18}$ Hz). The corresponding spectrum is expected to be flat, or ``red'', i.e. monotonically decreasing with the frequency (see Fig. \ref{fig5}).

Thanks to these properties,  a cosmic background of pre-big bang gravitational radiation is expected to be much  easier to detect than backgrounds produced by other models of inflation. Indeed, the presently operating, earth-based, interferometric gravitational antennas (such as LIGO, VIRGO, GEO, TAMA), and those planned to be operative in space (LISA, BBO, DECIGO), cover together a sensitivity band roughly ranging from the mHz frequency scale to the $10^2$ Hz scale (see e.g. \cite{26}). In this range  the expected signal of pre-big bang gravitational waves is possibly much higher than the primordial gravitational backgrounds originating from other sources. In Fig. \ref{fig5} we have explicitly plotted the sensitivity levels expected to be reached by the final, advanced versions of LIGO, VIRGO and BBO. We also recall that,  with the presently attained sensitivities, the limits imposed by earth-based interferometers on a stochastic background of comic gravitational waves (i.e.  $\Om_g h^2 \laq 10^{-6}$ at a frequency $\sim 10^2$ Hz, see e.g. \cite{limit}), is already better than the limit imposed by primordial nucleosynthesis.  Note that -- given the expected peak value of the pre-big bang spectrum-- such a limit already excludes, at high frequency, the possibility of a red spectrum, in agreement with the predictions of a generic model of pre-big bang inflation \cite{36a,36b}.

Conversely,  gravitational radiation of pre-big bang origin tends to be extremely  suppressed at the low frequency scales which are relevant for the observable properties of the CMB radiation. As a result, its contribution to CMB polarization effects \cite{37} is  negligible -- unless the growing spectrum of the pre-big bang gravitational waves is flat enough to keep a large intensity also at low frequencies, see e.g. \cite{pol}, as illustrated in Fig. \ref{fig5} (higher curvature corrections, in particular, may have a flattening effect on the spectrum \cite{68a}).

Thus, a combined non-observation of a primordial B-polarization mode in the CMB radiation, and a direct detection  of primordial high frequency gravitational waves, could represent a strong signal  in favor of the pre-big bang  scenario \cite{26}. In the cyclic/ekpyrotic scenario, on the contrary, the primordially produced background of cosmic gravitational waves is expected to be quite negligible today, both in the low and in the high-frequency range \cite{Boyle}. 

\subsubsection{\it{Scalar metric perturbations}}

The difficulty of a growing spectrum concerns, instead, the scalar part of the metric fluctuations. These fluctuations, amplified by inflation, should represent the primordial ``seeds" responsible for the peak structure of the temperature anisotropies $\Da T/T$, observed today in the CMB radiation: this requires, however, a nearly flat (or ``scale-invariant") spectral distribution, as confirmed by most recent observations \cite{5}. In models of pre-big bang inflation, on the contrary, the primordial scalar spectrum tends to be growing \cite{38} like the tensor spectrum, in striking disagreement with the data. 

In the absence of a brane-dominated regime, with an associated phase of slow-roll inflation able to produce the required spectrum of scalar perturbations, this problem can be solved by the so-called ``curvaton mechanism"  \cite{39,39a,39b}. Thanks to this mechanism, a  scale-invariant, adiabatic spectrum of scalar curvature perturbations can also be produced by the post-big bang decay of a suitable background field (other than the metric), whose scalar fluctuations have been amplified by inflation with a flat spectrum. 

In the context of the pre-big bang scenario the role of the curvaton could be played by the fundamental string theory axion  \cite{40}, associated by spacetime duality  to the four-dimensional components of the antisymmetric tensor $B_{\mu\nu}$ present in the string spectrum. The quantum fluctuations of the axion, in fact, can be amplified by a phase of pre-big bang inflation with a flat spectrum \cite{41,41a}, in sharp contrast with the quantum fluctuations of the metric tensor. 

Assuming that the axion fluctuations are amplified by a flat spectrum, it follows that  the  standard post-big bang phase, on large, super-horizon scales, is initially characterized by a sea of ``isocurvature" axion fluctuations, with a negligible component of  metric perturbations. These fluctuations may act as ``seeds" (i.e. as  quadratic sources to the perturbation equations) for the generation of isocurvature metric perturbations \cite{72a}, which are  allowed, but  only as a  possible sub-dominant component of the total observed anisotropy. However, 
if the post-big bang axion background acquires a mass and decays, 
the axion and metric fluctuations become linearly coupled. Thanks to this coupling new scalar (curvature) perturbations are automatically generated with the same spectral slope as the axion one, 
and with a spectral amplitude of the same order as the primordial axion amplitude \cite{39a,40}. The net result of this ``curvaton" mechanism is a (non-primordial) spectrum of adiabatic metric perturbations which, if sufficiently flat, may successfully reproduce the observed anisotropies of the CMB radiation. 

By applying the above mechanism, and assuming such an axion origin of the large-scale anisotropy, it becomes possible to extract from present CMB measurements direct information on models of pre-big bang inflation and on their fundamental parameters, such as 
the inflation scale in string units \cite{40}, or the kinematics of the extra dimensions. 

Let us recall, in fact, that the inflationary dynamics of the quantum fluctuations of the axion field $\sg$ is described by (the Fourier component of) the canonical variable $v_k(\eta)$, and is controlled, outside the horizon, by the so-called ``pump field'' $z(\eta)$. The pump field, in its turn, turns out to be determined by the cosmological background fields such as the background metric and the dilaton (or inflation) field. Let us suppose that, during inflation, the evolution of the pump field can be parametrized by a power-law behavior in terms of the conformal time coordinate $\eta$, i.e. $z(\eta) \sim |\eta|^\a$. In that case, the axion fluctuations $\sg_k= v_k/z$, after crossing the horizon $(k \eta \gaq 1)$, are characterized by a frozen (canonically normalized) spectral distribution given by
\beq
|\da_\sg|^2 = k^3 |\sg_k|^2 \sim k^{2+2\a} \equiv k^{n_s-1},
\label{18}
\eeq
where $n_s= 3+ 2 \a$ is the so-called scalar spectral index. 
A similar result is also valid for the metric perturbation spectrum, with the only difference that the power $\a$ is referred to the graviton pump field, different in general from the axion pump field even in the same given background. 

Consider now a  typical low-energy pre-big bang background, represented by (the negative-time branch of) a higher-dimensional vacuum solution of the tree-level gravi-dilaton equations. Let us suppose, for simplicity, that the background geometry can be described by a homogeneous Bianchi-I-type metric (namely by a cosmological metric which is homogeneous but not isotropic along all spatial directions), 
with three inflationary and isotropically expanding dimensions, with scale factor $a(\eta)$, and $n$ ``internal" dimensions, with scale factor $b_i(\eta)$, $i= 1, \dots , n$. In conformal time, and in the string frame, such solution can be parametrized for $\eta<0$ as \cite{42}
\beq
a \sim (-\eta)^{\b_0/(1-\b_0)}, ~~~~~
b_i \sim (-\eta)^{\b_i/(1-\b_0)}, ~~~~~
\phi \sim {\sum_i \b_i + 3 \b_0 -1\over 1-\b_0} \ln (-\eta),
\label{19}
\eeq
where the constant coefficients $\b_0,\b_i$ satisfy the Kasner-like condition
\beq
3 \b_0^2+ \sum_i \b_i^2 =1.
\label{20}
\eeq
The computation of the axion pump field then gives  \cite{41,41a}
\beq
z_\sg \sim a \left( \prod_i b_i\right)^{-1/2} e^{\phi/2} \sim (-\eta)^{(5\b_0-1)/ 2(1-\b_0)} \equiv (-\eta)^\a,
\label{22}
\eeq
and the corresponding spectral slope of the amplified axion fluctuations is:
\beq
n_s-1 \equiv 2+2\a= {1+3\b_0\over 1-\b_0}.
\label{23}
\eeq
The axion spectrum, in particular, turns out to be scale-invariant for $\b_0=-1/3$ (and $\sum_i \b_i^2=2/3$, according to Eq. (\ref{20})). Note that if $d=3+n=9$, as prescribed by critical superstring theory \cite{13,13a}, and if the external six-dimensional space is isotropic ($\b_i=\b$ for all directions), then the value of $\b_0$ required by a flat spectrum is automatically obtained if the internal dimensions expand or contract with the same kinematic power as the external ones, i.e. $|\b_i|=|\b_0|=1/3$. This seems to suggest a close (but still mysterious, to the best of our knowledge) connection between the isotropy of the internal and external background  and the scale-invariance of the spectrum. 

Let us now recall that, according to the 
most recent recent Planck results \cite{4},  the spectral slope of scalar perturbations (averaged over all scales) turns out to be
\beq
n_s \equiv 3+ 2\a= 0.9677 \pm 0.0060. 
\label{24}
\eeq
This slope slightly (but unambiguously) deviates from an exactly scale-invariant spectrum, corresponding to $n_s=1$ (or $2+2\a=0$). The comparison with Eq. (\ref{23}) then gives  $\b_0 \simeq -0.348$, which implies (using Eq. (\ref{20})) $\sum_i \b_i^2 \simeq 0.64$. A flat spectrum associated to a nine-dimensional background with the same (absolute) value of the internal and external kinematic power would give, instead,  $\b_0 =-1/\sqrt 9 \simeq -0.33$, and $\sum \b_i^2 =6/9\simeq 0.66$. The result (\ref{24}) may thus be interpreted, in $d=9$, as the indication of a small asymmetry between the absolute value of the power governing the time evolution of internal and external spatial dimensions. 

\subsubsection{\it{Primordial magnetic fields}}

We should add, finally, that the (number and) kinematics of the extra spatial dimensions, together with the dilaton kinematics, play a fundamental role also in the pre-big bang amplification of the quantum fluctuations of the electromagnetic fields, for a possible ``primordial" production of large-scale magnetic fields \cite{44a, 73a}. Remarkably, the values of $\b_0$ and $\b_i$ compatible with a (nearly) flat axion spectrum are also compatible with an efficient production of ``seeds" for the cosmic magnetic fields, for various types of photon-dilaton coupling (as is discussed in \cite{44b} for the specific examples of the  heterotic and type I superstring models). 

\section{Conclusion}

In this contribution we have briefly summarized the main features of a cosmological scenario proposed about twentyfive years ago \cite{11}, a scenario which naturally emerges from some very basic symmetries of string theory, which first suggested the possible interpretation of the big bang as a a curvature ``bounce", and which implements inflation as a phase of accelerated, growing curvature, growing coupling pre-big bang evolution. 

In such a context, the big bang is regarded not as a singular and unique event marking the beginning of everything (including space-time itself), but only as a transition (even if a dramatic one, in various respects) between two different (and possibly duality-related) cosmological regimes.

This pre-big bang scenario is able to to predict -- through the so-called curvaton mechanism based on the string theory axion -- a nearly flat, viable spectrum of scalar metric perturbations. The main observable difference from the standard (slow-roll) inflationary scenario is the associated production of primordial gravitational waves with a ``blue" (i.e. growing with frequency) spectral distribution. This peculiar prediction can be used to obtain, hopefully in the near future, a direct experimental disproof (or confirmation) of the pre-big bang scenario, or, more generally, to constrain the parameter space of pre-big bang models. 

We wish to conclude this short review with a comment concerning a possible ``cyclic" extension of the pre-big bang scenario, similar to that obtained in the  context of the ekpyrotic  \cite{43} and of the ``Conformal Cyclic" \cite{Pen} cosmological picture. A cyclic extension of the pre-big bang scenario could correspond, in particular, to a cyclic alternation of the phase of standard evolution and of its string-theory dual. However, as the dilaton is monotonically growing during the pre-big bang evolution, a cyclic scenario would require -- to close the cycle, and prepare the system to a new big bang transition -- a late-time  phase of decreasing dilaton, in order to lead back the system to the perturbative regime at the beginning of a new, self-dual cycle (see e.g. Fig. \ref{fig3}). 

Such a  bouncing back of the dilaton in the post-big bang regime could require the dominant contribution of a suitable (duality breaking? non-perturbative?) dilaton potential.  We thus naturally arrive at the question: could such a process of dilaton bouncing be associated to  the phase of (low-energy) cosmic acceleration that we are presently experiencing? See \cite{83,84,85} for a possible answer.

\acknowledgments
This work is supported in part by MIUR under grant no. 2012CPPYP7 
(PRIN 2012), and by INFN under the program TAsP ({Theoretical Astroparticle Physics}).


\begin{thebibliography}{0}
\newcommand{\bb}{\bibitem}




\bb{1}S. Weinberg,  {\em Gravitation and cosmology} (Wiley,
New York) 1971.

\bb{3}A. Guth, Phys. Rev. D {\bf 23}  (1981) 347.  

\bb{3a}V. F. Mukhanov, {\em Physical Foundations of Cosmology} (Cambridge University Press, Cambridge)  1981. 

\bb{5}Particle Data Group webpage at 
{\tt http://pdg.lbl.gov/} .

\bb{4}Planck Collaboration, {\em Planck
2015 results. I. Overview of products and scientific results}, 
arXiv:1502.01582 [astro-ph.CO].

\bb{6}A. D. Linde,  Phys. Lett. B {\bf 129}  (1983) 177. 

\bb{7}R. H. Brandenberger and J. Martin, Mod. Phys. Lett. A {\bf 16}  (2001) 999; Phys. Rev. D {\bf 63} (2001) 123501.

\bb{8}A. Vilenkin, Phys. Rev.  D {\bf 46}  (1992) 2355;  

\bb{8a}A. Borde and A. Vilenkin, Phys. Rev. Lett. {\bf 72} (1994)  3305; 

\bb{8b}A. Borde, A. Guth and A. Vilenkin, Phys. Rev. Lett. {\bf 90} (2003) 151301. 

\bb{9}V. Kaplunovsky, Phys. Rev. Lett. {\bf 55}  (1985) 1036.  

\bb{10}I. Antoniadis,  Phys. Lett. B {\bf 246}  (1990)  377.  

\bb{10a}N. Arkani-Hamed, S. Dimopoulos and G. R. Dvali,  Phys. Lett. B {\bf 429}  (1998) 263.  

\bb{11}M. Gasperini and  G. Veneziano, Astropart. Phys. 
{\bf 1}  (1993) 317;  Phys. Rep. {\bf 373} (2003) 1. 

\bb{12}G.  Veneziano, Phys. Lett. B {\bf 265} (1991).

\bb{13}M. B. Green, J. Schwarz and E. Witten, {\em Superstring theory} (Cambridge University Press, Cambridge) 1987.

\bb{13a}  
J. Polchinski, {\em String theory} (Cambridge University Press, Cambridge) 1998. 

\bb{blackhole} A. Strominger and C. Vafa, Phys. Lett. B {\bf 379}  (1996) 99.

\bb{stringcollision}D. Amati, M. Ciafaloni and G. Veneziano, Phys. Lett. B {\bf 216}  (1989) 41. 

\bb{temp}G. Veneziano, Europhys. Lett. {\bf 2}  (1986) 199. 

\bb{temp1}L. Susskind,  {\em Some speculations about black hole entropy in string theory}, hep-th/9309145.

\bb{13b} K. Kikkawa and M. Y. Yamasaki, Phys. Lett. B  {\bf 149}  (1984)  357.   

\bb{14}B. A. Campbell, A. Linde and K. A. Olive, Nucl. Phys. B {\bf 355}  (1991) 146. 

\bb{14a}R. Brustein and P. J. Steinhardt, Phys. Lett. B {\bf 302}  (1993) 196.  

\bb{15}C. Burgess et al, JHEP {\bf 0107}  (2001) 047. 

\bb{17}F. Quevedo, Class. Quantum Grav. {\bf 19}  (2002) 5721. 

\bb{18}J. Khoury et al, Phys. Rev.  D {\bf 64}  (2001) 123522.  

\bb{18a}J. Khoury et al, Phys. Rev.  D {\bf 65}  (2002) 086007.  

\bb{burt}E. I. Buchbinder, J. Khoury and B. A. Ovrut, 
Phys. Rev. D {\bf 76} (2007) 123503. 

\bb{Axel}K. Becker, M. Becker and A. Krause, Nucl. Phys. B {\bf 715}  (2005) 349. 

\bb{Liddle} A. R. Liddle, A. Mazumdar and F. E. Schunk, Phys. Rev. D {\bf 58}  (1998) 061301.

\bb{20}S. Kachru et al, JCAP {\bf 0310}  (2003) 013. 

\bb{21}S. Kachru et al,  Phys. Rev.  D {\bf 68}  (2003) 0460005. 

\bb{22}S. Giddings, S. Kachru and J. Polchinski, Phys. Rev.  D {\bf 66}  (2002) 106006. 

\bb{31}A. Buonanno, T. Damour and G. Veneziano, 
Nucl. Phys. B {\bf 543}  (1999) 275. 


\bb{23}A. A. Tseytlin, Mod. Phys. Lett. A {\bf 6}  (1991) 1721. 

\bb{24}K. A. Meissner and G. Veneziano, Mod. Phys. Lett. A 
{\bf 6}  (1991) 3397; Phys. Lett. B {\bf 267}  (1991) 33.
  
\bb{25}M. Gasperini and G. Veneziano,  Phys. Lett. B {\bf 277}  (1992)
256. 

\bibitem{17a} 
  R. Brustein, M. Gasperini and  G. Veneziano, 
  Phys.\ Lett.\ B {\bf 431} (1998) 277.

\bb{26}M. Gasperini, {\em Elements of string cosmology}
(Cambridge University Press, Cambridge)  2007.

\bb{26a}R. Brustein and G. Veneziano, Phys. Lett. B {\bf 329}  (1994)
429. 
 
\bb{27}M. Gasperini, M. Maggiore and G. Veneziano, Nucl. Phys. B {\bf 494}  (1997) 315.  

\bb{27a}R. Brustein and R. Madden, Phys. Rev. D. {\bf 57}  (1998)
712. 

\bb{27b}C. Cartier, E. J. Copeland and R. Madden, JHEP {\bf 0001}  (2000) 035. 

\bb{28}M. Gasperini, M. Giovannini and G. Veneziano, Phys. Lett. B  {\bf 569}  (2003) 113; Nucl. Phys. B {\bf 694}  (2004) 206.   
   
\bb{29}M. Gasperini, M. Giovannini, K. A. Meissner and G.
Veneziano, Nucl. Phys. (Proc. Suppl.) B {\bf 49}  (1996) 70-74.  

\bibitem{20a} 
  M. Gasperini and M. Giovannini,
  Class.\ Quant.\ Grav.\  {\bf 10} (1993) L133. 

\bb{42}M. Gasperini and G. Veneziano, Phys. Rev. D {\bf 50}  (1994) 2519.   

\bb{30}N. Kaloper, A. Linde and R. Bousso, Phys. Rev. D {\bf 59}  (1999) 043508. 


\bb{32} M. Gasperini, Phys. Rev. D {\bf 61}  (2000) 
87301. 

\bb{33}M. Gasperini, Classical Quant. Grav. {\bf 17}  (2000) R1.

\bb{33a}M. Gasperini, Mod. Phys. Lett. A {\bf 14}  (1999) 1059.  

\bb{34}R. Brandenberger and C. Vafa, Nucl. Phys. B {\bf 316}  (1999) 391. 

\bb{34a}A. A. Tseytlin and C. Vafa, Nucl. Phys. B {\bf 372}  (1992) 443. 

\bb{53a}G. Veneziano, JCAP {\bf 0403}  (2004)  004.  

\bb{34b}S. Alexander, R. Brandenberger and D. Easson, {Phys. Rev.} D {\bf 62}  (2000) 103509. 

\bb{Wat}T. Battenfeld and S. Watson, Rev. Mod. Phys. {\bf 78}  (2006) 435. 

\bb{35}M. Gasperini and G. Veneziano, Gen. Rel. Grav.
{\bf 28}  (1996) 1301.  

\bb{35a}M. Gasperini, J. Maharana and G. Veneziano, Nucl. Phys. B {\bf 472} (1996) 349.   

\bb{35b}M. Gasperini,  Int. J. Mod. Phys. D {\bf 10}  (2001) 15.  

\bb{36}M. Gasperini and M. Giovannini, )Phys. Lett. B {\bf 282}  (1992) 
36.  

\bb{36a}M. Gasperini and  M. Giovannini, Phys. Rev.  D {\bf 47}  (1993) 1519. 

\bb{36b}R. Brustein, M. Gasperini, M. Giovannini and  G.
Veneziano, Phys. Lett. B {\bf 361}  (1995) 45.  

\bb{36c}R. Brustein, M. Gasperini and  G. Veneziano, Phys.  Rev. D 
{\bf  55}  (1997) 3882.  



\bb{Bicep}BICEP2/Keck and Planck Collaborations, {\em A joint analysis of BCEP2/Keck Array and Planck dat}, arXiv:1502.00612 [astro-ph.CO].

\bb{limit} J. Aasi et al [LIGO and VIRGO Collaboration], Phys. Rev. Lett. {\bf 113}  (2014) 231101.  

\bb{37}M. Kamionkowski, A. Kosowski and A. Stebbins, Phys. Rev. Lett. {\bf 78}  (1997) 2058.  

\bb{pol}M. Gasperini, {\em Elementary introduction to pre - big bang cosmology and to the relic graviton background}, in {\em Gravitational waves}, edited by I. Ciufolini, V. Gorini, U. Moschella and P. Fr\`e (Taylor \& Francis) 2001, p. 280-337 [hep-th/9907067].

\bb{68a}
 M.~Gasperini,
  Phys.\ Rev.\ D {\bf 56} (1997) 4815.

\bb{Boyle}L. A. Boyle, P. J. Steinhardt and N. Turok, Phys. Rev. D {\bf 69}  (2004) 127302. 

\bb{38}R. Brustein, M. Gasperini, M. Giovannini,
V. Mukhanov and  G. Veneziano, Phys. Rev. D {\bf 51}  (1995) 6744.

\bb{39}K. Enqvist and M. Sloth, Nucl. Phys. B {\bf 626}  (2002)  395. 
 
\bb{39a}D. H. Lyth and D. Wands, Phys. Lett. B {\bf 524}  (2002) 5.   

\bb{39b}T. Moroi and T. Takahashi, Phys. Lett. B {\bf 522}   (2001) 215.

\bb{40}V. Bozza, M. Gasperini, M. Giovannini and G. Veneziano, Phys. Lett.  B {\bf 543} (2002)  14;   Phys. Rev. D {\bf 67} (2003) 063514.
 
\bb{41}E. J. Copeland, R. Easther and  D. Wands,  Phys. Rev. 
 {\bf D56} (1997) 874.  
  
\bb{41a}E. J. Copeland, J. E. Lidsey and D. Wands, Nucl. Phys. B {\bf 506}  (1997) 407.  

\bb{72a}R. Durrer, M. Gasperini, M. Sakellariadou and G. Veneziano,  Phys. Rev. D {\bf 59} (1999) 043511;  
  Phys.\ Lett.\ B {\bf 436} (1998) 66 
  [astro-ph/9806015].

\bb{44a}M. Gasperini, M. Giovannini and G. Veneziano, Phys. Rev. Lett. {\bf 75}   (1995) 3796;  Phys. Rev. 
D {\bf 52} (1995) 665.   

\bb{73a} D. Lemoine and M. Lemoine, Phys. Rev. D {\bf 52}  (1995) 1995.    

\bb{44b}M. Gasperini and S. Nicotri, Phys. Lett. B {\bf 633}  (2006) 155.  

\bb{43}P. J. Steinhardt and N. Turok, Phys. Rev. D {\bf 65}  (2002) 126003;  Science {\bf 296} (2002)1436.  

\bb{Pen}R. Penrose, {\em Cycles of time: an extraordinary new view of the Universe} (Bodley Head, London) 2010.

\bb{83} M.~Gasperini, F.~Piazza and G.~Veneziano,
  Phys.\ Rev.\ D {\bf 65} (2002) 023508.

\bb{84}M.~Gasperini,
  Phys.\ Rev.\ D {\bf 64}  (2001) 043510.

\bb{85}
L.~Amendola, M.~Gasperini and F.~Piazza,
  JCAP {\bf 0409} (2004) 014 .

\end{thebibliography}
\end{document}